\documentclass[a4paper]{doublecol-new}

\usepackage{natbib}
\usepackage{stfloats}
\usepackage{mathrsfs}
\usepackage{graphicx}
\usepackage{fixltx2e}
\usepackage{amsmath,amssymb,url}
\usepackage{placeins}
\usepackage{hyperref}
\usepackage{multirow}

\usepackage[table,xcdraw]{xcolor}

\begin{document}%


\LRH{I. Kazemian and S. Aref}

\RRH{A Minimax Regret Model for the Capacitated Hub Location Problem with Multiple Allocations}








\title{Hub Location under Uncertainty: a Minimax Regret Model for the Capacitated Problem with Multiple Allocations}

\authorA{Iman Kazemian}
\affA{Department of Industrial Engineering\\College of Engineering\\University of Tehran, North Kargar\\Tehran, PO box: 4563-11155, Iran\\
        Tel: +98-913-1965513\\E-mail: i.kazemian@ut.ac.ir}

\authorB{Samin Aref}
\affB{Department of Computer Science\\University of Auckland\\Auckland, Private Bag 92019, New Zealand\\
Department of Computing and Information Technology\\Unitec Institute of Technology\\Auckland, Private Bag 92025, New Zealand\\
 E-mail: sare618@aucklanduni.ac.nz}

\begin{abstract}

In this paper the capacitated hub location problem is formulated by a minimax regret model, which takes into account uncertain setup cost and demand. We focus on hub location with multiple allocations as a strategic problem requiring one definite solution. Investigating how deterministic models may lead to sub-optimal solutions, we provide an efficient formulation method for the problem.

A computational analysis is performed to investigate the impact of uncertainty on the location of hubs. The suggested model is also compared with an alternative method, seasonal optimization, in terms of efficiency and practicability. The results indicate the importance of incorporating stochasticity and variability of parameters in solving practical hub location problems. Applying our method to a case study derived from an industrial food production company, we solve a logistical problem involving seasonal demand and uncertainty. The solution yields a definite hub network configuration to be implemented throughout the planning horizon.
\end{abstract}

\KEYWORD{Hub location, Uncertain demand, Uncertain setup cost, Capacitated, Multiple allocations, Minimax Regret model, Case Study, Robust optimization, Re-configuration Cost, Seasonal demand.}


\begin{bio}

Iman Kazemian holds a B.S. degree in Industrial Engineering from Isfahan University of Technology as well as an M.S. degree from Sharif University of Technology. He is an Industrial Engineering doctoral candidate at University of Tehran. His research interests include Mathematical Programming, Operations Management, and Logistics.

Samin Aref holds an M.S. degree in Industrial Engineering from Sharif University of Technology. He is a Computer Science doctoral candidate at University of Auckland and a lecturer at Unitec Institute of Technology. His fields of interest include Complex Social and Economic Systems, Operations Research, and Structural Analysis of Social Networks.

Authors have presented at New Zealand Mathematics Colloquium and Conferences of Iranian Operations Research Society. They have published in \textit{Global Journal of Flexible Systems Management}, \textit{International Journal of Occupational Safety and Ergonomics}, \textit{European Journal of Operational Research}, and \textit{Journal of Revenue \& Pricing Management}.

\textbf{Reference} to this paper should be made as follows: Kazemian, I. and Aref, S. (2017) ‘Hub Location under Uncertainty: a Minimax Regret Model for the Capacitated Problem with Multiple Allocations’, Int. J. Supply Chain and Inventory Management, Vol. 2, No. 1, pp.1-19. \url{https://doi.org/10.1504/IJSCIM.2017.086371}

\end{bio}

\maketitle

\clearpage

\section{Introduction}
\label{sec:1}

Hub networks are one of the most common types of logistics systems serving urban transportation, airline networks, communication systems, and cargo networks. The basic characteristic of hub networks is that routing is performed through a subset of the connections between nodes instead of direct connections from the origin to the destination. The communication industry seems to be the first platform for using hubs, while decades has passed since hubs networks were institutionalized in logistics systems, transportation industry, air cargoes, and postal services. Nowadays, hub network design is a common practice for wholesalers, distribution companies, and food production industries whose main objective is to enhance logistics efficiency. Hub network configuration suggests using a set of hubs and spokes for connecting different origins and destinations. Different industries make use of hubs to deal with logistics activities in a more productive way by reducing direct transportation paths. Drawing an analogy might be helpful to clarify the importance of hub networks. A complete directed graph with $K$ nodes has $K\times(K-1)$ arcs, while all the nodes can be connected to each other by having a central node (hub) being connected to all the other peripheral nodes (spokes) which reduces the number of arcs to $2\times(K-1)$. That is how the connectivity is achievable by utilizing fewer resources more productively. Hub location problem (HLP) originates from this idea; the challenge of deciding on allocation of hubs to obtain an efficient logistics network.

The main objective in a common HLP is to minimize the total costs of establishing hubs and transportation of products between hubs and spokes. Hub location problems are categorized into capacitated and uncapacitated problems such that the former embodies most of the real-world problems. After HLP was introduced, subsequent problems like p-median, p-hub center, and hub covering problems were emerged to address different locational challenges of the industry.

The principal purpose of p-median is to locate a number of hubs in the network so that the total transportation cost is minimized. The second problem, p-hub center, aims to optimize location of the hubs and allocation of nodes such that the cost of major routes in the network is minimized. In the hub covering problem, minimizing the total cost by finding the optimal location of the nodes and their corresponding allocation shapes the question where the number of hubs is not predefined. Such a problem introduces limits of coverage as the number of nodes that are connected to a hub is limited.

Equally relevant to the problem type, main objective, and the decision variables are the questions of single and multiple network allocation patterns i.e. whether the spokes are to be connected to one hub or multiple hubs.

\section{Literature Review}
\label{sec:1.5}

\begin{figure*}[ht]
  \includegraphics[width=\textwidth]{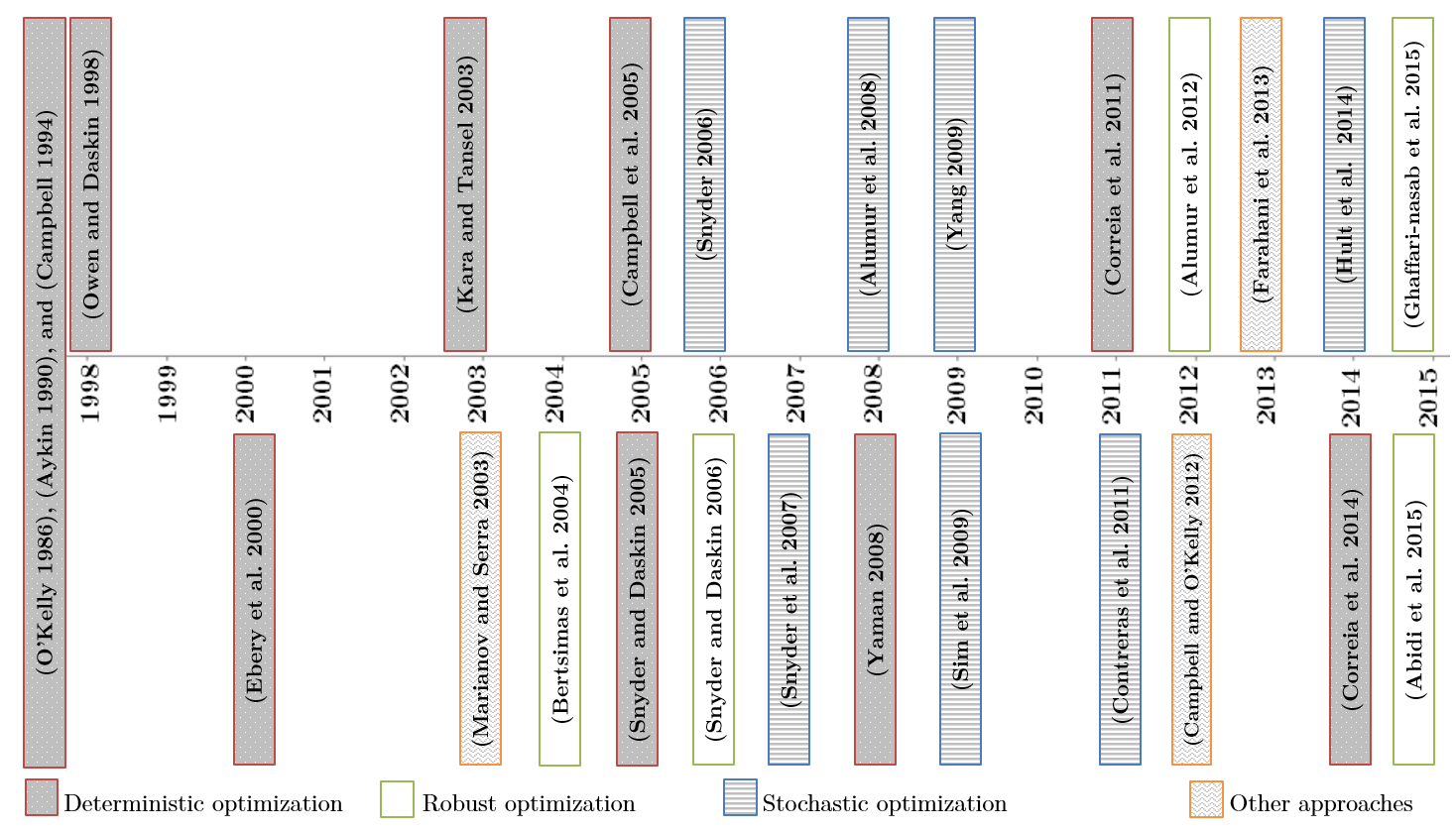}
\caption{The evolution of optimization methods used for hub location problem: a transition from deterministic models to stochastic and robust models}
\label{fig:0}       
\end{figure*}

HLP was first posed \citet{okelly86}. The author introduced Single-HLP concerning assignment of the appropriate location to the hub and its connection to the spokes in a setting where there is no cost for hub establishment and it had infinite capacity. He then used a mathematical model to formalize another location problem on hub airline network \citet{okelly87}. The basic model made progress towards P-HLP, a quadratic model, incorporating a number of hubs with direct transportation routes \citet{aykin1990quadratic}. The proceeding linear model was extended to p-median location problem \citet{campbell90} to incorporate multiple network allocation patterns. A more comprehensive mathematical model for multiple allocation HLP was presented \citet{campbell94} embodying real world assumptions such as fixed cost for connecting spokes to hubs, minimum flow, and the capacity of nodes. Although most of the hub location models developed assumed potential locations for the hubs in a discrete space, there were early research studies relaxing this assumption and considering a continuous space \citet{okelly86active,aykin92}.

As the logistics operations got more complicated, new problems emerged with different objective functions, formulations, and solution techniques. 
The taxonomy of HLP suggests four ramifications of the original problems including Capacitated p-Median Problem \citet{shamsi}, HLP with star network structure \citet{yaman2008,yaman2012star}, p-hub center problem \citet{campbell2005hub,campbell2005hub2}, and p-hub covering problem \citet{kara2003single}.
For a more detailed review of the literature one may refer to \citet{alumur2008network,farahani2013hub}.

In the pioneering research studies the parameters were assumed to be deterministic though it was not realistic. Like the other location and logistics problems, incorporation of stochastic parameters is a promising research area that is receiving increasing attention. Some recent and more realistic approaches towards the model definition are as follows. \citet{marianov2003location} developed a model for locating hubs in the network of air transportation by formulating a M/D/c queuing system . The same research field was investigated by \citet{yang2009stochastic} who developed a two-stage stochastic programming model for air transportation with uncertainty in demand . They introduced a stochastic programming model to address the uncapacitated air freight hub location and flight routes planning under seasonal demand variations \citet{yang2009stochastic}.

Sim, Lowe, and Thomas, suggested a single assignment hub covering model where the arc travel time was normally distributed with given mean and standard deviation. The objective was to locate p hubs in order to minimize the longest transportation time in the network for a specified service level in delivery times. In their formulation, they did not include the cost of establishing the hubs and operating the network \citet{sim2009stochastic}. Moreover, Contreras et al. designed a model for multiple allocation HLP with uncertainty in both demands and transportation costs. They proved that stochastic problems with uncertain demands or dependent transportation costs are equivalent to a deterministic problem in which random variables are replaced by their expected values. However, in the case of uncertain independent transportation cost, the corresponding stochastic problem is not equivalent to the such a deterministic model \citet{contreras2011stochastic}. 
Snyder et al. presented a robustness measure that combines the two objectives by minimizing the expected cost while bounding the relative regret in each scenario for two classical facility location problems: the k-median problem and the uncapacitated fixed-charge location problem \citet{snyder2006stochastic}. In particular, their models seek a minimum expected cost solution that is p-robust.

More recently, Alumur presented multiple allocation and single allocation HLP considering uncertainty factors in demand and setup costs \citet{alumur2012hub}. Abidi and Razmi suggested the same model as \citet{alumur2012hub} (stochastic uncapacitated hub location problem with multiple allocations) for a similar case study where setup cost uncertainty was replaced by travel cost uncertainty \citet{adibi20152}. Some other approaches were suggested to move towards the reality in hub location network design. Risk pooling and its effects on hub location network design is suggested by \citet{snyder2007stochastic}. Reliability models for facility location are another approach taking the expected failure cost into account \citet{snyder2005reliability}. Hult, Jiang, and Ralph solved the uncapacitated single allocation p-hub center problem with stochastic travel times using cutting planes and Benders decomposition \citet{hult2014exact}.  For a complete review on facility location problems with different sources of uncertainty, see \citet{campbell2012twenty}. Figure \ref{fig:0} demonstrates the evolution of methods used for solving hub location problems through time. A transition from deterministic optimization method to more sophisticated methods like robust and stochastic optimization is evident.

\begin{table*}[ht]
\caption{Basic model notation}
\label{tab:1}      
\begin{tabular}{ l l}
\hline\noalign{\smallskip}
$N = \{ 1,2,...,n \} $ & Set of nodes \\
$ F_k $ & Fixed setup cost $ k \in N $ \\
$ d_{ij} $ & Distance from node $i \in N$ to node $j \in N$ \\
$ \beta $ & Collection cost \\
$ \delta $ & Distribution cost \\
$ \alpha $ & Transfer cost \\
$ W_{ij} $ & Flow routed from node $ i \in N $ to node $j \in N$ \\
$ C_{ijkm} $ & The transportation cost from node $i \in N$ to node $j \in N$ routed via the hubs $k \in N$ and $m \in N$\\
$ \Gamma_k $ & The capacity of node $k \in N$ if a hub is located there\\
$x_{ijkm}$ & Fraction of flow from node $i \in N$ to node $j \in N$ routed via the hubs $k \in N$ and $m \in N$\\
$y_k$ &  Hub or spoke assignment for node $k \in N$\\
\noalign{\smallskip}\hline
\end{tabular}
\end{table*}

As stated earlier hub location is an essential part of the strategic planning for distribution companies having far-reaching effects on their operational issues and productivity. Besides, as logistics activities are changing within time, the data used in hub network decisions can become outdated by the network utilization phase. Therefore, some of the parameters required for designing the network cannot be determined accurately. The most common uncertain parameters are costs, distances, and demands. Failure to consider uncertainty of parameters may lead to obtaining sub-optimal network designs as the determinant input parameters change. The cause of uncertainty of such parameters are as follows. The volatility of costs for initial procurement such as land, industrial equipment, and construction makes the setup cost uncertain. Although, the demand can be predicted by market research, the lag time between designing the network and its actual utilization makes many predictions outdated especially for the case of time-dependent demands like seasonal products. This issue indicates that uncertainty should also be considered for demand parameters. In many cases uncertain parameters follow a familiar probability distribution which needs stochastic optimization. There are also cases where the data does not fit to familiar distributions which require robust programming to take uncertainty into consideration. In both situations considering different scenarios in a discrete probabilistic space would take uncertainty into account \citet{alumur2012hub}.

\subsection*{Our contribution:} \label{ss:contrib}
This research aims to investigate the effect of uncertainty on the solutions obtained from different modeling techniques proposed by the contemporary researchers. It concerns different approaches toward the formulation of a typical HLP with comprehensive uncertainty factors and capacity constraints. In many real world problems capacity constraints are an indispensable component of hub location mathematical modeling.

However, to the best of our knowledge there was no published research studies on the investigation of uncertainty impacts on real hub location problems with capacity constraints and inaccurate decision parameters. This research study fulfills a needed gap on capacitated HLP under uncertainty as suggested for future research by many recent research studies \citet{alumur2012hub,adibi20152,ghaffari2015robust}. In this study, a novel mathematical model is proposed to question this hypothesis whether deterministic modeling and seasonal optimization can be sound measures for obtaining the optimal location of the hubs under different sources of uncertainty.

The structure of rest of the paper is as follows. Section \ref{sec:2} presents a deterministic model to introduce the model foundation. Section \ref{sec:3} suggests a more sophisticated optimization model with uncertain parameters. The efficiency of optimization model is evaluated in the Section \ref{sec:4} by analyzing a numerical example. Finally, a practical case is discussed in Section \ref{sec:5} to demonstrate the impact of the proposed approach in solving real world problems.

\section{Basic Model}
\label{sec:2}

\begin{table*}[ht]
\caption{Minimax regret model notation}
\label{tab:2}       
\begin{tabular}{l l}
\hline\noalign{\smallskip}
$S'_f $ & All the scenarios with different uncertain setup costs $ s' \in S'_f$ \\
$ F_{s'}^k $ & The cost of establishing a hub at node $k$ in scenario $ s'$ \\
$S_w$ & All the scenarios with different uncertain demands $s \in S_w$ \\
$P_s$ & The probability that scenario $s \in S_w$ occurs\\
$W_{ij}^s$ & Demand routed from node $i$ to node $j$ in scenario $s \in S_w$ \\
\noalign{\smallskip}\hline
\end{tabular}
\end{table*}

A review on the current literature reveals the presence of research studies on HLP with capacity constraints \citet{ebery2000capacitated,correia2011hub,correia2014multi}. However, the capacitated hub location literature requires realistic approaches towards incorporating uncertainty in modeling such as minimax regret model. This study suggests a novel modeling approach with four linear constraints to deal with uncertainty in demand and setup costs. As mentioned earlier, the foundation of optimization model is first introduced by replicating a deterministic model originally developed by \citet{campbell94} and then the main approach of modeling is outlined in the next section to analyze capacitated HLP with multiple allocations. Adapting from a deterministic model proposed in a well-known study by \citet{campbell94}, the basic model notation is stated in Table \ref{tab:1}. The total cost is structured in a way that it incorporates hubs setup cost as well as three types of transportation costs including collection cost for factory to hub, transfer cost for hub to hub, and distribution cost for hub-to-spoke. Accordingly, the transportation cost is formulated in Eq. \eqref{eq:1}.

\begin{equation}\label{eq:1}
  C_{ijkm}=\beta d_{ik} + \alpha  d_{km}  + \delta d_{mj} \qquad  \forall  i,j,k,m \in N
\end{equation}
The capacitated HLP with multiple allocations is formulated as follows in \eqref{eq:2} to \eqref{eq:7}.
\begin{equation}\label{eq:2}
  \min \sum\limits_{i} \sum\limits_{j} \sum\limits_{k} \sum\limits_{m} W_{ij} C_{ijkm} x_{ijkm} + \sum\limits_{k}F_k y_k
\end{equation}
S.t.
\begin{equation}\label{eq:3}
  \sum\limits_{m}x_{ijkm} \leq y_{k}  \qquad  \forall  i,j,k \in N
\end{equation}
\begin{equation}\label{eq:4}
  \sum\limits_{k} x_{ijkm}\leq y_m  \qquad  \forall  i,j,m \in N
\end{equation}
\begin{equation}\label{eq:5}
    \sum\limits_{i}{\sum\limits_{j}} W_{ij} \sum\limits_{m} x_{ijkm} \leq \Gamma_{k} y_{k}  \qquad  \forall k \in N
\end{equation}
\begin{equation}\label{eq:6}
  y_k \in\{0,1\}  \qquad  \forall k \in N
\end{equation}
\begin{equation}\label{eq:7}
  0 \leq x_{ijkm}\leq 1  \qquad  \forall  i,j,k,m \in N
\end{equation}
In this formulation, $x_{ijkm}$ stands for the decision variable which is the fraction of flow from location $i$ (origin) to location $j$ (destination) that is routed via hubs at locations $k$ and $m$ in that order. $y_k$ represents a binary variable showing whether node $k$ is a hub by taking one, or it is a spoke by taking zero. Eq. \eqref{eq:3} and \eqref{eq:4} ensures that the flow passes through the hubs. Eq. \eqref{eq:5} takes the capacity of the hubs into account. Finally, domain constraints are formulated in \eqref{eq:6} and \eqref{eq:7}.

\section{Minimax Regret Model}
\label{sec:3}
The impact of hub location on total cost makes it a crucial strategic decision to be made in logistics. As an attempt to deal with lack of precise information on the operational parameters of logistics networks, minimax regret model can be deployed. Minimax regret is a robust optimization approach to minimize the worst-case regret. The aim of this technique is to perform as closely as possible to the optimal course. Since the minimax criterion applied here is to the regret rather than to the payoff itself, it is not as pessimistic as the common minimax approach.

One benefit of minimax regret models are that they are independent to the probabilities of the various outcomes. Morover, in comparison with other methods of robust optimization, it is easy to use specially in cases where regret can be accurately computed while probabilities of scenarios are difficult to estimate. Although minimization of maximum regret is a common practice, many other techniques have been proposed for both stochastic and robust optimization in HLP. Ghaffari-Nasab et al. considered the capacitated single and multiple allocation HLP with stochastic demands \citet{ghaffari2015robust}. The main emphasis of their paper relies on robust optimization tools developed by \citet{bertsimas2004price} for linear programming problems. Situation and strategies in logistics decision making need different approaches to tackle the uncertainty.  For a review of the application of such approaches to facility location problems, see \citet{owen1998strategic} and \citet{snyder2006facility}.

It is assumed in this paper that the uncertainty in demand can be described by considering a limited number of scenarios. It is also assumed that in each scenario demand parameters are certain values. Moreover, uncertain behavior of setup costs is assumed to be interpretable by considering different scenarios. Deploying such scenarios alongside minimax regret programming, the model will be able to tackle real world problems in an uncertain environment \citet{alumur2012hub}. The minimax regret model notation as originally suggested by \citet{alumur2012hub} for another type of HLP is stated in Table \ref{tab:2}.

\begin{table*}[ht]
\caption{Test case 1: different solutions for capacitated HLP with multiple allocations}
\label{tab:3}       
\begin{tabular}{p{1cm} llllllll} \hline
~ & \multicolumn{2}{p{1in}}{$\alpha $=0.3} & \multicolumn{2}{p{1in}}{$\alpha $=0.5} & \multicolumn{2}{p{1in}}{$\alpha $=0.7} & \multicolumn{2}{p{1in}}{$\alpha $=1} \\ \cline{2-9}
 & Cost & Hub & Cost  & Hub & Cost  & Hub & Cost  & Hub \\ \cline{2-9}
BDM & 2905117 & 2,3 & 2989450 & 1,3 & 3065952 & 1,3 & 3138530 & 1,3 \\
$s_{f1}$ & 2884970 & 2,3 & 2969830 & 2,3 & 3054288 & 2,3 & 3138440 & 2,3 \\
$s_{f2}$ & 2084747 & 2,3 & 2169607 & 2,3 & 2254065 & 2,3 & 2338217 & 2,3 \\
$s_{f3}$ & 2461068 & 2,4 & 2547230 & 2,4 & 2630818 & 2,4 & 2712427 & 2,4 \\
$s_{f4}$ & 1779440 & 1,3 & 1864380 & 1,3 & 1942708 & 1,3 & 2018130 & 1,3 \\
MRM & - & 3,4 & - & 3,4 & - & 3,4 & - & 3,4 \\ \hline
\end{tabular}
\end{table*}

\begin{equation}\label{eq:8}
\begin{aligned}
  Z_{s'}^*= \min \sum\limits_{s}P_s \sum\limits_{i} \sum\limits_{j} \sum\limits_{k} \sum\limits_{m} W_{ij}^s C_{ijkm} x_{ijkm} \\ 
	+ \sum\limits_{k}F_{s'}^k  y_k
\end{aligned}
\end{equation}
S.t.
\begin{equation}\label{eq:9}
  \sum\limits_{m}x_{ijkm} \leq y_{k}  \qquad  \forall  i,j,k \in N
\end{equation}
\begin{equation}\label{eq:10}
  \sum\limits_{k} x_{ijkm}\leq y_m  \qquad  \forall  i,j,m \in N
\end{equation}
\begin{equation}\label{eq:11}
    \sum\limits_{i}{\sum\limits_{j}} W^s_{ij} \sum\limits_{m} x_{ijkm} \leq \Gamma_{k} y_{k}  \qquad  \forall k ,s \in N
\end{equation}
\begin{equation}\label{eq:12}
  y_k \in\{0,1\}  \qquad  \forall k \in N
\end{equation}
\begin{equation}\label{eq:13}
  0 \leq x_{ijkm}\leq 1  \qquad  \forall  i,j,k,m \in N
\end{equation}
$Z_{s'}^*$  is the optimal solution of the model above. Respectively as the exact scenario that will occur is not known, a minmax regret model can be considered as in Eq. \ref{eq:14} to Eq. \ref{eq:20} in which the maximum regret is to be minimized. In this mathematical model, the representation of $y_k$ is the same as before.  The above model can be easily linearized by defining variable $R$ such that $R \geq R_s' \quad \forall \quad s' \in S'_f$.
\begin{equation}
\label{eq:14}
 \min \max _{s' \in S'_f} = R_s'
\end{equation}
S.t.
\begin{equation}\label{eq:15}
  \sum\limits_{m}x_{ijkm} \leq y_{k}  \qquad  \forall  i,j,k \in N
\end{equation}
\begin{equation}\label{eq:16}
  \sum\limits_{k} x_{ijkm}\leq y_m  \qquad  \forall  i,j,m \in N
\end{equation}
\begin{equation}\label{eq:17}
    \sum\limits_{i}{\sum\limits_{j}} W^s_{ij} \sum\limits_{m} x_{ijkm} \leq \Gamma_{k} y_{k}  \qquad  \forall k ,s \in N
\end{equation}
\begin{equation}\label{eq:18}
\begin{aligned}
R_s'=\sum\limits_{s}P_s \sum\limits_{i} \sum\limits_{j} \sum\limits_{k} \sum\limits_{m} W_{ij}^s C_{ijkm} x_{ijkm} + \\
\sum\limits_{k}F_{s'}^k  y_k-Z_{s'}^* \qquad  \forall s' \in S'_f
\end{aligned}
\end{equation}
\begin{equation}\label{eq:19}
  y_k \in\{0,1\}  \qquad  \forall k \in N
\end{equation}
\begin{equation}\label{eq:20}
  0 \leq x_{ijkm}\leq 1  \qquad  \forall  i,j,k,m \in N
\end{equation}

\section{Computational Analysis}
\label{sec:4}
To test the model proposed in Section \ref{sec:3}, some test problems associated with the distribution of products among five different cities \ref{tabap1}-- \ref{tabap3} were used (highlighted cells to be discussed later). The other parameters were considered as $\beta=\delta=1$  and  $\alpha \in \{0.3,0.5,0.7,1\}$. Four different scenarios for uncertainty in the setup cost were designed in which $F_{s'}^k$ was randomly selected. Moreover, for analyzing the uncertainty in demands, four different scenarios were selected with equal probabilities $P_s=0.25$. GAMS software was used to solve the numerical example.

Optimal solutions of different modes are represented in Table \ref{tab:3}.
The basic deterministic model is abbreviated to BDM and is shown in the first row. The four scenarios of the stochastic model are presented in the next rows according to four different values assumed for cost of land represented by $s_f$. Finally, minimax regret model is abbreviated to MRM and is demonstrated in the lowest row. The problems are solved separately based on each scenario considering minmax regret models. The solutions can be compared with that obtained from the basic deterministic model in which setup costs and demands were set to the mean values.

The second column form left side of Table \ref{tab:3} shows the total cost and the third column represents the optimal location of the hubs. Note that as the objective function in the minmax regret model is represented differently, total cost cannot be compared with that of scenarios. Hence, the cells for values of the minimax regret cost are left empty. As was evident in Table \ref{tab:3}, the optimum location of the hubs in the minimax regret model differs from the other scenarios. It can be concluded from the observation that it is more appropriate to use a minimax regret model instead of estimating costs and demands or using a deterministic scenario.

No relationship is observed between the costs of setting up a hub and selecting a location for the hub. For example, node $4$ has the highest setup cost, but in some problems it is selected as the optimal location of hub. This observation indicates that in addition to the setup cost, the demand and geographical location are also determinant factors in the optimal solution of capacitated HLP.

In order to show the robustness of the model to re-designing the position of the points, a new city is added to the test case. The highlighted rows and columns in Tables \ref{tabap1} -- \ref{tabap3} show the added city distances and cost under different scenarios as well as demand from the other cities. The modified problem is solved using GAMS. As expected, the optimal location of hubs for MRM would not change from cities 3 and 4 which shows the robustness of minimax regret model to points of sale. Furthermore, the minimax regret model is robust to the variation of demand and cost as already demonstrated in Table \ref{tab:3}.

As an attempt to investigate larger test cases, randomly generated problems with 5,6,10 and 15 nodes are analyzed in which $\beta=\delta=1$  and  $\alpha=0.5$. For each problem, 4 scenarios of setup cost and 4 scenarios of demand (with equal probabilities $P_s=0.25$) are considered. The results are reported in Table \ref{tab:3.2}. As it is illustrated in Table \ref{tab:3.2} the optimal location of hubs can be changed based on the modeling method (deterministic versus minimax regret) and variations in demand. This supports the previous argument that change of parameters requires re-configuration of hub network if the model is deterministic. However, taking the variations of parameters into account, a minimax regret model can be used to determine a definite hub network configuration that works well under uncertainty. This approach is implemented on a practical case in the next Section.
\begin{table}[ht]
\centering
\caption{The optimal location of hubs for the other test cases}
\label{tab:3.2}
\begin{tabular}{lllll}
\hline
Test case    & 2   & 3   & 4   & 5    \\ \hline
No. of nodes & 5   & 6   & 10  & 15   \\
BDM          & 1,2 & 1,3 & 5,3 & 5,2  \\
$s_{f1}$     & 3,4 & 1,3 & 3,9 & 5,11 \\
$s_{f2}$     & 1,4 & 1,3 & 1,5 & 2,4  \\
$s_{f3}$     & 1,3 & 1,4 & 2,3 & 1,5  \\
$s_{f4}$     & 1,3 & 1,4 & 1,9 & 3,13 \\
MRM          & 1,3 & 1,3 & 5,9 & 1,12 \\ \hline
\end{tabular}
\end{table}

\clearpage
\section{Practical Case Study}
\label{sec:5}
The application of proposed model is evaluated using the data from an Iranian industrial food production company. The case of chocolate production in Shirin Asal Tabriz Co. is a well-known local hub location example, investigated by \citet{rostami}. They analyzed the impact of estimated demand on the configuration of hub networks in different scenarios. According to their model, the location of hubs can be changed seasonally. In contrast, they did not consider the costs of seasonal re-configuration the network as the model required so. As already discussed, we consider HLP as a strategic decision making process requiring an unchangeable solution as proposed by this research.

As mentioned earlier, the fluctuation of prices makes hub establishment an uncertain activity in terms of monetary issues. Moreover, according to the data reported by the company, the demand for products is seasonal and the setup cost is highly dependent on time, making it an appropriate case to be analyzed by our model due to the fluctuations in demand and variability of the setup costs. The scenarios are designed by dividing the year into four seasons with equal length for demand and considering five scenarios for setup costs with $0.7F_k$ to $1.3F_k$.

The geographical structure of Shirin Asal Tabriz Co. market is as follows. The main factory is located in the city of Tabriz in the north-west of Iran supporting 36 distribution points with demand across the country. The national market can be divided into three regions namely: the west, the center and the east. The data presented in this study are related to the demand of the west part of the country with 14 nodes. The company management sought to establish hubs among these cities (14 locations). Table \ref{tab:4} outlines the demands in each scenario. Table \ref{tab:5} includes the capacities and the cost (in million Rials) of establishing hubs in each location. The problem is to find the best assignment of the hubs to the cities and their allocation according to nodes’ capacities and uncertain demand and setup cost. According to the studies performed for this particular example, $\alpha = 0.4, \beta = 1$ and $\delta = 1$ are calculated \citet{rostami} and $P_s=0.25$ to represent four seasons. The optimization is performed by GAMS software on a personal computer with 2.4 GHz Intel Core i5 CPU and 4 GB of RAM.

\begin{table}[ht]
\caption{Demands in different seasons}
\label{tab:4}       
\begin{tabular}{lrrrr} \hline
City & Spring & Summer & Fall & Winter \\ \hline
Rasht & 9205 & 7899 & 21848 & 26510 \\
Kermanshah & 10459 & 11256 & 18751 & 19628 \\
Tabriz & 48022 & 39529 & 65890 & 94831 \\
Tehran & 14412 & 12571 & 44070 & 91906 \\
Zanjan & 10590 & 11732 & 20402 & 21218 \\
Qazvin & 4424 & 5995 & 10848 & 12502 \\
Hamedan & 6270 & 4802 & 9273 & 9505 \\
Urmia & 17022 & 16006 & 24951 & 23234 \\
Ardabil & 13764 & 19839 & 19281 & 16767 \\
Sanandaj & 7996 & 6105 & 11330 & 10807 \\
Shahrekord & 6142 & 5721 & 10320 & 9065 \\
Ilam & 4044 & 4135 & 6689 & 7881 \\
Karaj & 18519 & 22050 & 41018 & 48957 \\
Arak & 4272 & 3726 & 9402 & 10287 \\  \hline
\end{tabular}
\end{table}

\begin{table}[ht]
\centering
\caption{Capacity and setup cost for different cities}
\label{tab:5}   
\begin{tabular}{lllll}
\hline
\multirow{2}{*}{City} & \multirow{2}{*}{Capacity} & \multicolumn{3}{c}{Setup cost} \\ \cline{3-5} 
                      &                           & Min      & Avg.      & Max     \\ \hline
Rasht & 275200       &910.56&  1300.80& 1691.04\\
Kermanshah & 154200  &431.76&  616.80 & 801.84\\
Tabriz & 321640      &900.59&  1286.56& 1672.52\\
Tehran & 38528       &1358.78& 1941.12& 2523.45\\
Zanjan & 201360      &703.80&  1005.44& 1307.07\\
Qazvin & 123840     &346.75&  495.36&  643.96\\
Hamedan & 97600      &329.28&  470.40&  611.52\\
Urmia & 110080       &308.22&  440.32&  572.41\\
Ardabil & 123840     &346.75&  495.36&  643.96\\
Sanandaj & 103200    &288.96&  412.80&  536.64\\
Shahrekord & 137600 &385.28&  550.40&  715.52\\
Ilam & 82560         &231.16&  330.24&  429.31\\
Karaj & 460960       &1500.68& 2143.84& 2786.99\\
Arak & 82835         &231.93&  331.34&  430.74\\  \hline
\end{tabular}
\end{table}

\begin{figure}[ht]
  \includegraphics[totalheight=0.7\textheight]{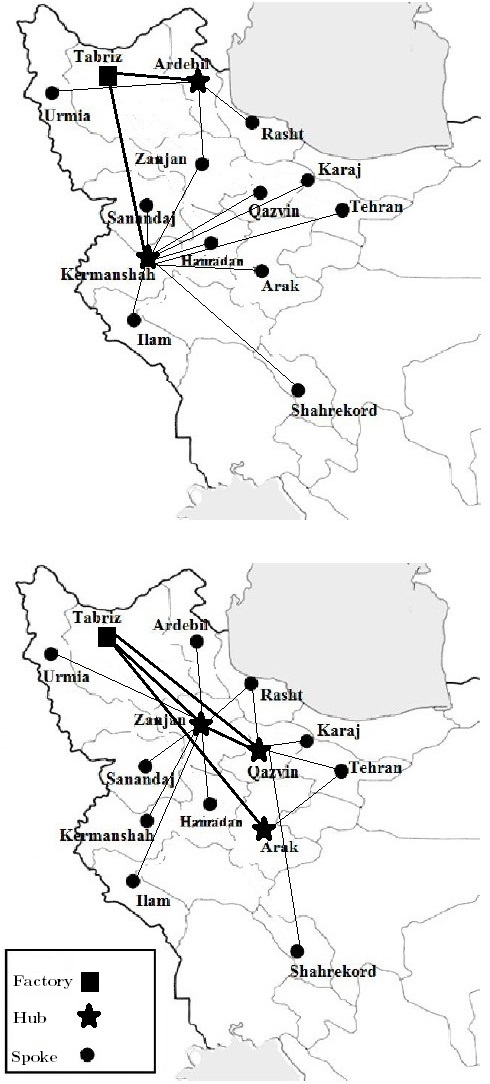}
\caption{Different mathematically optimal hub location solutions obtained from deterministic model (top) minimax regret model (bottom)}
\label{fig:1}       
\end{figure}

To demonstrate the practical application of the proposed minimax regret model, a deterministic model (setup costs and demands set to the mean values) is first solved and then the minimax regret model is used to analyze the case with respect to significant sources of uncertainties discussed. The result of deterministic model and minimax regret model are shown in Fig. \ref{fig:1} where the flow between the factory and the hubs (cost represented by $\beta$) as well as that between the hubs (cost represented by $\alpha$) are represented by thick lines and the flow between hubs and spokes (cost represented by $\delta$) is represented by narrow lines. The network obtained from the deterministic model is shown on the map on top of Fig. \ref{fig:1} suggesting establishing two hubs in cities of Ardebil and Kermanshah, basing the decision on mean values and ignoring the uncertainty. On the other hand, the minimax regret model suggests establishing hubs in three cities of Qazvin, Zanjan and Arak as the solution is illustrated on the map in the bottom of Fig. \ref{fig:1}.

It is noteworthy that in the optimal solution obtained by the minimax regret model, two hubs located in cities of Qazvin and Zanjan are connected. Such solutions are not likely be optimal in an uncapacitated setting. However, in capacitated problems if high demand nodes are close to low capacity nodes with low setup cost, we may observe the optimal solution connecting hubs. In this case, the connectivity of hubs in Qazvin and Zanjan is mainly due to the high demand in cities of Tehran and Karaj and relatively low capacity of Qazvin as their close hub with low setup cost.

Furthermore, the most striking observations to emerge from the network configuration are the distribution routes from the company to cities of Ardebil and Urmia. As evident in Fig. \ref{fig:1} the products need to be first collected in Zanjan (hub) and then be distributed to the two cities of Ardebil and Urmia.  Although based on the distances such an optimal solution might seem odd, the hub establishment cost values render this reasonable. Direct transportation of products from the factory requires Tabriz to be a hub, while Tabriz is the fourth most expensive locations for hub establishment. Worthy of mention is that the factory cannot support the functions of the hub and for the purposes of the problem Tabriz is similar to the other cities.

\begin{figure}[ht]
  \includegraphics[width=0.48\textwidth]{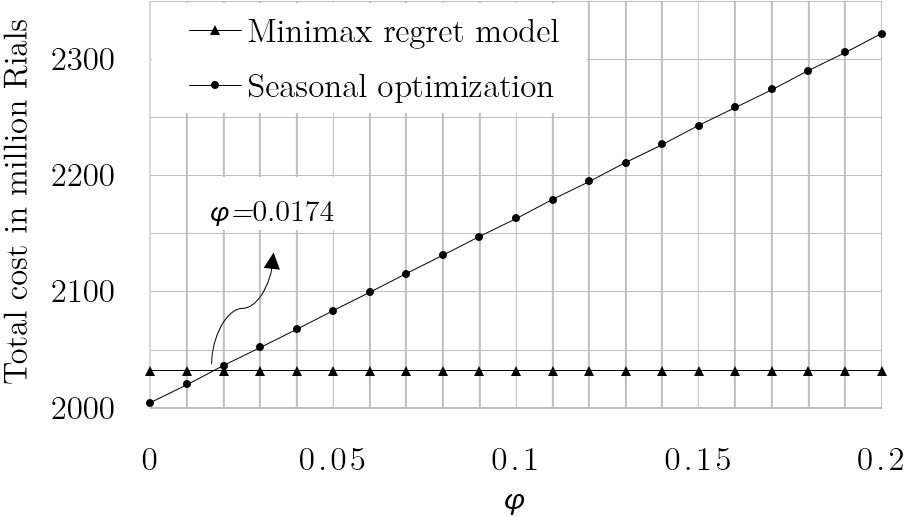}
\caption{Comparison of total cost from seasonal optimization and minimax regret model based on different values of $\phi$}
\label{fig:2}       
\end{figure}

The total cost of the network obtained by seasonal optimization model is computed a 960 day operation time \citet{rostami}. The reported cost is exclusive of network seasonal re-configuration cost involved in operations dealing with training, human resources, and facility transportation. If we consider $\phi$ as the ratio of re-configuration cost (compromising of operations dealing with training, human resources, and facility transportation) to the total cost of hub network establishment, results from the two models can be compared as in Fig. \ref{fig:2} showing a break-even point for 1 year operation. As evident in the figure, our suggested model is capable of reducing the total cost for $\phi \geq 0.0174$ and the cost reduction is proportional to the ratio defined $\phi$. Therefore, for industries where hub network is easy to be re-configured every season based on demand variations, seasonal optimization suggested by \citet{rostami} can be used. For the other cases where the company needs to spend a non-trivial ($\phi \geq 0.0174$ in here) amount of money on hub network re-configurations, a definite hub network design as suggested in the present study is more cost-effective.

The substantial difference between the solutions obtained from the deterministic model and minimax regret model shows the impact of uncertainty on HLP. The differences in hub network designs let us argue that deterministic analysis of HLP suffers from capability of industrial practice in cases where there are source of uncertainty. Deterministic analysis may lead to sub-optimal solutions imposing excessive costs to the company in the long term.  Therefore, deploying a definite solution obtained from a comprehensive model is suggested to deal with logistics with seasonal demand and uncertainty to be implemented throughout the planning horizon.

\clearpage
\section{Application Perspectives}
Designing logistics network under uncertainty has applications already in demand in various businesses across different industries including those with a seasonal demand and other time-variant decision parameters. This study aims to broaden the hub location body of knowledge via a quantitative investigation of an industrial problem to assist executives by providing a novel network design model. The suggested model is to be used for increasing cost-effectiveness as a competitive advantage as well as meeting other operational objectives.

In more specific terms, this study suggests a method to tackle uncertainty in locational decision making to meet essential technical challenges associated with designing hub networks like inaccuracy of estimates and variability of demand. A minimax regret model can be used in decision making problems where the uncertainty can be modeled by considering a number of scenarios for the uncertain parameter while no historical data is available to estimate probability distributions required for stochastic programming. Practical examples provided show how the suggested technique can be used to obtain a definite logistics network under uncertain demand and setup cost.

\section{Conclusion}
\label{sec:6}
In this paper, multiple allocation capacitated HLP was investigated in a setting where setup cost and demand were both uncertain. After outlining a well-know model in Section \ref{sec:2}, different extensions are discussed and a minimax regret model was developed in Section \ref{sec:3} for considering such sources of uncertainty. A computational analysis was performed in Section \ref{sec:4} to investigate the changes in the optimal location of the hub caused by different modeling approaches. The result showed that the optimal solution changes where the model is associated with uncertain parameters. According to the numerical example in Section \ref{sec:5}, ignoring the uncertainty may change the whole hub location solution drastically. Moreover, the industrial application of the proposed method was addressed to by discussing a case from an industrial food production company. The results obtained from solving the minimax regret model for the case study and comparing it to the alternative modeling technique confirmed the efficiency of our suggested model in providing a definite solution for an industrial-sized problem challenged by an uncertain environment.

There are a number of research directions that are hoped to be investigated in the near future. Firstly, one may introduce a more sophisticated problem with pervasive sources of uncertainty by parameters not necessarily following a familiar distribution function to be solved by stochastic programming and compared with the technique used in this paper. It would contribute to our idea if the research is associated with a critical industrial-sized case of an uncertain environment and counterproductive practices.  Secondly, it is suggested for further research to tackle large HLPs of this type by tailored evolutionary algorithms and analyze practical cases for different datasets such as international couriers.


\appendix
\section{Computational analysis and case study data}

\begin{table}[ht]
\centering
\caption{Distance between 5 cities used in the computational analysis example}
\label{tabap1}
\begin{tabular}{llllll}
\hline
City & 1 & 2   & 3   & 4   & 5   \\ \hline
1 & 0 & 590 & 485 & 325 & 348 \\
2 &   & 0   & 588 & 526 & 414 \\
3 &   &     & 0   & 599 & 280 \\
4 &   &     &     & 0   & 319 \\
5 &   &     &     &     & 0   \\ 
\rowcolor[HTML]{EFEFEF} 
6 & 729 & 456 & 578 & 423 & 253 \\ \hline
\end{tabular}
\end{table}

\begin{table}[ht]
\centering
\caption{Four scenarios of demand used in the computational analysis example}
\label{tabap2}
\begin{tabular}{llllll}
\hline
\multicolumn{6}{l}{Demand scenario 1}        \\ \hline
City & 1     & 2     & 3     & 4     & 5     \\ \hline
1    &       & 12612 & 18085 & 16652 & 63467 \\
2    & 86249 &       & 92152 & 42150 & 6979  \\
3    & 15295 & 53550 &       & 87155 & 81086 \\
4    & 86940 & 90960 & 47946 &       & 88818 \\
5    & 61337 & 91633 & 76636 & 91141 &       \\ 
\rowcolor[HTML]{EFEFEF} 
6    & 45400 & 54222 & 35200 & 48576 & 43253 \\ \hline
\multicolumn{6}{l}{Demand scenario 2}        \\ \hline
City & 1     & 2     & 3     & 4     & 5     \\ \hline
1    &       & 68051 & 78747 & 43698 & 48346 \\
2    & 71429 &       & 67029 & 38488 & 44321 \\
3    & 39460 & 28955 &       & 73469 & 62608 \\
4    & 63443 & 7932  & 90296 &       & 68353 \\
5    & 19322 & 12575 & 6864  & 20751 &       \\ 
\rowcolor[HTML]{EFEFEF} 
6    & 54378 & 68782 & 42200 & 78245 & 65200 \\ \hline
\multicolumn{6}{l}{Demand scenario 3}        \\ \hline
City & 1     & 2     & 3     & 4     & 5     \\ \hline
1    &       & 49129 & 72170 & 91123 & 80320 \\
2    & 65650 &       & 26966 & 53580 & 26892 \\
3    & 63409 & 34737 &       & 16355 & 77912 \\
4    & 18540 & 57047 & 67416 &       & 25912 \\
5    & 14567 & 24116 & 84892 & 27186 &       \\ 
\rowcolor[HTML]{EFEFEF} 
6    & 86245 & 9237  & 83240 & 9257  & 8424 \\ \hline
\multicolumn{6}{l}{Demand scenario 4}        \\ \hline
City & 1     & 2     & 3     & 4     & 5     \\ \hline
1    &       & 35764 & 29767 & 10636 & 15561 \\
2    & 21637 &       & 72711 & 8641  & 55549 \\
3    & 26601 & 57047 &       & 52084 & 46490 \\
4    & 59851 & 53809 & 38386 &       & 4810  \\
5    & 46845 & 87287 & 55457 & 88819 &       \\ 
\rowcolor[HTML]{EFEFEF} 
6    & 16784 & 17865 & 12232 & 18526 & 15423 \\ \hline
\end{tabular}
\end{table}

\begin{table*}[ht]
\caption{Capacities and four scenarios of setup cost used in the computational analysis example}
\label{tabap3}
\begin{tabular}{llllll
>{\columncolor[HTML]{EFEFEF}}l }
\hline
City                     & 1          & 2          & 3          & 4          & 5          & 6          \\ \hline
Capacity                 & 682423     & 765892     & 876543     & 986578     & 546879     & 454786     \\
Setup cost in scenario 1 & 1414016725 & 965058131  & 1483010032 & 1529540551 & 1607592118 & 1825643271 \\
Setup cost in scenario 2 & 1213461250 & 727806327  & 920038779  & 1826350935 & 818597258  & 1934765489 \\
Setup cost in scenario 3 & 1710445940 & 1329737654 & 1936001128 & 720329416  & 1203047686 & 1745980390 \\
Setup cost in scenario 4 & 758042396  & 1890660867 & 622937195  & 1643029427 & 1699171001 & 1693049714 \\
Average setup cost       & 1273991578 & 1228315745 & 1240496784 & 1429812582 & 1332102016 & 1799859716 \\ \hline
\end{tabular}
\end{table*}

\begin{table*}[ht]
\caption{The distance between 14 cities of the practical case study}
\label{tabap4}
\begin{tabular}{lllllllllllllll}
\hline 
               & 1 & 2   & 3   & 4   & 5   & 6   & 7   & 8   & 9   & 10  & 11   & 12  & 13  & 14  \\ \hline
1. Rasht       & 0 & 590 & 485 & 325 & 348 & 185 & 401 & 739 & 266 & 565 & 868  & 774 & 285 & 577 \\
2. Kermanshah  &   & 0   & 588 & 526 & 414 & 433 & 189 & 582 & 791 & 136 & 731  & 184 & 538 & 365 \\
3. Tabriz      &   &     & 0   & 599 & 280 & 455 & 609 & 308 & 219 & 452 & 1142 & 772 & 574 & 785 \\
4. Tehran      &   &     &     & 0   & 319 & 150 & 337 & 907 & 591 & 501 & 543  & 710 & 50  & 239 \\
5. Zanjan      &   &     &     &     & 0   & 175 & 329 & 588 & 377 & 278 & 862  & 598 & 282 & 505 \\
6. Qazvin      &   &     &     &     &     & 0   & 244 & 763 & 451 & 453 & 584  & 617 & 106 & 303 \\
7. Hamedan     &   &     &     &     &     &     & 0   & 610 & 667 & 164 & 568  & 373 & 354 & 176 \\
8. Urmia       &   &     &     &     &     &     &     & 0   & 527 & 446 & 1178 & 766 & 729 & 786 \\
9. Ardabil     &   &     &     &     &     &     &     &     & 0   & 655 & 1134 & 975 & 552 & 843 \\
10. Sanandaj   &   &     &     &     &     &     &     &     &     & 0   & 732  & 320 & 523 & 340 \\
11. Shahrekord &   &     &     &     &     &     &     &     &     &     & 0    & 719 & 579 & 392 \\
12. Ilam       &   &     &     &     &     &     &     &     &     &     &      & 0   & 706 & 514 \\
13. Karaj      &   &     &     &     &     &     &     &     &     &     &      &     & 0   & 322 \\
14. Arak       &   &     &     &     &     &     &     &     &     &     &      &     &     & 0   \\ \hline
\end{tabular}
\end{table*}

\bibliographystyle{agsm}
\bibliography{science}

\end{document}